\documentclass[12pt]{article}
\usepackage{showkeys}
\begin{document}
\title{Planck Oscillators and Elementary Particles}
\author{B.G.Sidharth\\
Dipartimento di Matematica e Informatica,\\ Universita di Udine,
Via delle Scienze 206,\\ 33100 Udine Italy}
\date{}
\maketitle
\begin{abstract}
In this paper we argue that it is possible to model the Universe
along thermodynamic lines, using oscillators in the background Dark
Energy at the Planck scale. This leads to meaningful results in
agreement with observation.
\end{abstract}
\section{Introduction}
In 1997 the author proposed a model in which elementary particles
are created out of a background Dark Energy or Zero Point Field
(ZPF). This predicted a cosmology in which the universe accelerates,
driven by the Dark Energy, with a small cosmological constant.
Moreover several supposedly inexplicable so called Large Number
coincidences were all deduced from the above model
(Cf.refs.\cite{mg8,ijmpa,ijtp}). At that time the prevailing model
was the standard Big Bang in which, because of dark matter, the
universe would be decelerating. However the work of Perlmutter and
others confirm the cosmic acceleration and the small cosmological
constant in 1998 itself (Cf. for example \cite{perl,cu,uof} and
several references therein). Moreover in the new model there would
be a residual energy that would be extremely small.\\
Let us first re-derive the recently discovered \cite{mersini}
residual cosmic energy directly from the background Dark Energy. We
may reiterate that the ``mysterious'' background Dark Energy is the
same as the quantum Zero Point Fluctuations in the background vacuum
electromagnetic field .The background Zero Point Field is a
collection of ground state oscillators \cite{mwt}. The probability
amplitude is
$$\psi (x) = \left(\frac{m\omega}{\pi \hbar}\right)^{1/4} e^{-(m\omega/2\hbar)x^2}$$
for displacement by the distance $x$ from its position of classical
equilibrium. So the oscillator fluctuates over an interval
$$\Delta x \sim (\hbar/m\omega)^{1/2}$$
The background \index{electromagnetic}electromagnetic field is an
infinite collection of independent oscillators, with amplitudes
$X_1,X_2$ etc. The probability for the various oscillators to have
amplitudes $X_1, X_2$ and so on is the product of individual
oscillator amplitudes:
$$\psi (X_1,X_2,\cdots ) = exp [-(X^2_1 + X^2_2 + \cdots)]$$
wherein there would be a suitable normalization factor. This
expression gives the probability amplitude $\psi$ for a
configuration $B (x,y,z)$ of the magnetic field that is described by
the Fourier coefficients $X_1,X_2,\cdots$ or directly in terms of
the magnetic field configuration itself by, as is known,
$$\psi (B(x,y,z)) = P exp \left(-\int \int \frac{\bf{B}(x_1)\cdot \bf{B}(x_2)}{16\pi^3\hbar cr^2_{12}} d^3x_1 d^3x_2\right).$$
$P$ being a normalization factor. At this stage, we are thinking in
terms of energy without differenciation, that is, without
considering Electromagnetism or Gravitation etc as separate. Let us
consider a configuration where the field is everywhere zero except
in a region of dimension $l$, where it is of the order of $\sim
\Delta B$. The probability amplitude for this configuration would be
proportional to
$$\exp [-((\Delta B)^2 l^4/\hbar c)]$$
So the energy of \index{fluctuation}fluctuation in a region of
length $l$ is given by finally, the density \cite{mwt,uof}
$$B^2 \sim \frac{\hbar c}{l^4}$$
So the energy content in a region of volume $l^3$ is given by
\begin{equation}
\beta^2 \sim \hbar c/l\label{4e1}
\end{equation}
This energy is minimum when $l$ is maximum. Let us take $l$ to be
the radius of the Universe $\sim 10^{28}cms$. The minimum energy
residue of the background Dark Energy or Zero Point Field (ZPF)
alluded to now comes out to be $10^{-33}eV$, exactly the observed
value. This observed residual energy is a cosmic footprint of the
ubiquitous Dark Energy in the Universe, a puzzling footprint that,
as we have noted, has recently been observed \cite{mersini}. The
minimum mass $\sim 10^{-33}eV$ or $10^{-65}gms$, will be seen to be
the mass of the photon. Interestingly, this also is the minimum
thermodynamic mass in the Universe, as shown by Landsberg from a
totally different point of view, that of thermodynamics \cite{land}.\\
If on the other hand we take for $l$ in (\ref{4e1}) the smallest
possible length, which has been taken to the Planck length $l_P$, as
we will see in
the sequel, then we get the Planck mass $m_P \sim 10^{-5}gm$.\\
So (\ref{4e1}) gives two extreme masses, the Planck mass and the
photon mass. We will see how it is possible to recover the
intermediate elementary particle mass from
these considerations later.\\
As an alternative derivation, it is interesting to derive a model
based on the theory of phonons which are quanta of sound waves in a
macroscopic body \cite{huang}. Phonons are a mathematical analogue
of the quanta of the electromagnetic field, which are the photons,
that emerge when this field is expressed as a sum of Harmonic
oscillators. This situation is carried over to the theory of solids
which are made up of atoms that are arranged in a crystal lattice
and can be approximated by a sum of Harmonic oscillators
representing the normal modes of lattice oscillations. In this
theory, as is well known the phonons have a maximum frequency
$\omega_m$ which is given by
\begin{equation}
\omega_m = c \left(\frac{6\pi^2}{v}\right)^{1/3}\label{4e2}
\end{equation}
In (\ref{4e2}) $c$ represents the velocity of sound in the specific
case of photons, while $v = V/N$, where $V$ denotes the volume and
$N$ the number of atoms. In this model we write
$$l \equiv \left(\frac{4}{3} \pi v\right)^{1/3}$$
$l$ being the inter particle distance. Thus (\ref{4e2}) now becomes
\begin{equation}
\omega_m = c/l\label{4e3}
\end{equation}
Let us now liberate the above analysis from the immediate scenario
of atoms at lattice points and quantized sound waves due to the
Harmonic oscillations and look upon it as a general set of Harmonic
oscillators as above. Then we can see that (\ref{4e3}) and
(\ref{4e1}) are identical as
$$\omega = \frac{mc^2}{\hbar}$$
So we again recover with suitable limits the extremes of the Planck
mass and the photon mass.\\
Other intermediate elementary particle masses follow if we take $l$
as a typical Compton wavelength.The Compton wavelength comes about,
if we consider the entire Universe which has a volume of the order
$10^{84}$ cc. to be made up of $10^{120}$ Planck Oscillators,as we
will see.Then $v$ in (\ref{4e2}) turns out to be $10^{-36}$
and $l$ becomes $10^{-12}$ cm -- a typical Compton wavelength.\\
Max Planck had noticed that, what we call the Planck scale today,
$$l = \left(\frac{\hbar G}{c^3}\right)^{\frac{1}{2}} \sim
10^{-33}cm$$
$$m = \sqrt{\frac{\hbar c}{G}} \sim 10^{-5}gm$$
\begin{equation}
t = \sqrt{\frac{\hbar G}{c^5}} \sim 10^{-42} sec\label{4ea1}
\end{equation}
is made up of the fundamental constants of nature and so, he
suspected it played the role of a fundamental length \cite{kiefer}.
Indeed, modern Quantum Gravity approaches have invoked (\ref{4ea1})
in their quest for a reconciliation of gravitation with other
fundamental interactions. Indeed, as can be seen from (\ref{4ea1}),
this scale combines the Gravitational constant $G$ and classical
theory with the Planck constant of Quantum theory. However if this
is a fundamental scale, then the time honoured prescription of a
differentiable spacetime has to be abandoned.\\
There is also another scale made up of fundamental constants of
nature, viz., the well known Compton scale (or classical electron
radius),
\begin{equation}
l = e^2/m_ec^2 \sim 10^{-12}cm\label{4ea2}
\end{equation}
where $e$ is the electron charge and $m_e$ the electron mass. The
Compton scale emerges from the ZPF. This had appeared in the
Classical theory of the electron unlike the Planck scale, which was
a product of Quantum Theory. Indeed if (\ref{4ea2}) is substituted
for
$l$ in (\ref{4e1}), we get the elementary particle mass scale.\\
The scale (\ref{4ea2}) has also played an important role in modern
physics, though it is not considered as fundamental as the Planck
scale. Nevertheless, the Compton scale (\ref{4ea2}) is close to
reality in the sense of experiment, unlike (\ref{4ea1}), which is
well beyond foreseeable direct experimental contact. Moreover
another interesting feature of the Compton scale is that,
it brings out the Quantum Mechanical spin, unlike the Planck scale.\\
Let us investigate further from another point of view.
\section{The Planck and Compton Scales}
 String Theory, Loop Quantum Gravity and a few
other approaches start from the Planck scale. This is also the
starting point in our alternative theory of Planck oscillators in
the background dark energy. We first give a rationale for the fact
that the Planck scale would be a minimum scale in the Universe
\cite{bgsijmpe}. Our starting point \cite{bgsfpl172004,uof} is the
model for the underpinning at the Planck scale for the Universe.
This is a collection of $N$ Planck scale oscillators where we will
specify $N$ shortly.\\ Let us consider an array of $N$ particles,
spaced a distance $\Delta x$ apart, which behave like oscillators
that are connected by springs. We then have
\cite{bgsfpl152002,good,vandam,uof}
\begin{equation}
r  = \sqrt{N \Delta x^2}\label{4De1d}
\end{equation}
\begin{equation}
ka^2 \equiv k \Delta x^2 = \frac{1}{2}  k_B T\label{4De2d}
\end{equation}
where $k_B$ is the Boltzmann constant, $T$ the temperature, $r$ the
extent  and $k$ is the spring constant given by
\begin{equation}
\omega_0^2 = \frac{k}{m}\label{4De3d}
\end{equation}
\begin{equation}
\omega = \left(\frac{k}{m}a^2\right)^{\frac{1}{2}} \frac{1}{r} =
\omega_0 \frac{a}{r}\label{4De4d}
\end{equation}
We now identify the particles with \index{Planck}Planck
\index{mass}masses and set $\Delta x \equiv a = l_P$, the
\index{Planck}Planck length. It may be immediately observed that use
of (\ref{4De3d}) and (\ref{4De2d}) gives
$$k_B T \sim m_P c^2,$$
which of course agrees with the temperature of a \index{black
hole}black hole of \index{Planck}Planck \index{mass}mass. Indeed,
Rosen \cite{rosen} had shown that a \index{Planck}Planck
\index{mass}mass particle at the \index{Planck scale}Planck scale
can be considered to be a \index{Universe}Universe in itself with a
Schwarzchild radius equalling the Planck length. We now observe that
if in (\ref{4De1d}), we take the extent $r$ to be the radius of the
universe $\sim 10^{27}cm$, then we get for $N$, the number of Planck
oscillators underpinning the universe, $N \sim 10^{120}$, which we
had assumed earlier. If we now use the well known fact that there
are $10^{80}$ elementary particles in the universe then we conclude
that $n_r \sim 10^{40}$ Planck oscillators underpin elementary
particles \cite{bgsfpl172004,bgsfpl152002}. In any case, ultimately,
the only parameter we will be using is the well known number of
elementary
particles in the universe $\sim 10^{80}$.\\
Using this in (\ref{4De1d}), we get $r \sim l$, the \index{Compton
wavelength}Compton wavelength $10^{-12}cm$ as required. Whence the
elementary particle mass is given by
\begin{equation}
m = m_P/\sqrt{n} \sim 10^{-25}gm\label{e10}
\end{equation}
This follows from (\ref{4De4d}), remembering that $\omega =
mc^2/\hbar$. This shows that while the Planck frequency or energy is
the highest, the energy of the $n \sim 10^{40}$ Planck oscillator
array is lowest, so that this configuration is stable. In fact we
recover from (\ref{e10}), the Compton wavelength (\ref{4ea2}) and
Compton time of an elementary particle.\\
This explains why we encounter \index{elementary
particles}elementary particles, rather than \index{Planck}Planck
\index{mass}mass particles in nature. In fact as known \cite{bhtd},
a \index{Planck}Planck \index{mass}mass particle decays via the
\index{Bekenstein radiation}Bekenstein radiation within a
\index{Planck time}Planck time $\sim 10^{-42}secs$. On the other
hand, the lifetime of an elementary particle
would be very much higher.\\
We now make two interesting comments. Cercignani and co-workers have
shown \cite{cer1,cer2} that when the gravitational energy becomes of
the order of the electromagnetic energy in the case of the Zero
Point oscillators, that is
\begin{equation}
\frac{G\hbar^2 \omega^3}{c^5} \sim \hbar \omega\label{4e5}
\end{equation}
then this defines a threshold frequency $\omega_{max}$ above which
the oscillations become chaotic. In other words, for meaningful
physics we require that
$$\omega \leq \omega_{max}.$$
Secondly as we saw from the parallel but unrelated theory of phonons
\cite{huang,reif}, which are also bosonic oscillators, we deduce a
maximal frequency given by
\begin{equation}
\omega^2_{max} = \frac{c^2}{l^2}\label{4e6}
\end{equation}
In (\ref{4e6}) $c$ is, as we saw in the particular case of phonons,
the velocity of propagation, that is the velocity of sound, whereas
in our case this velocity is that of light. Frequencies greater than
$\omega_{max}$ in (\ref{4e6}) are again meaningless. We can easily
verify that using (\ref{4e5}) in (\ref{4e6}) gives
\begin{equation}
Gm_P^2 \sim \hbar c \sim \epsilon^2\label{ex}
\end{equation}
This is a well known relation expressing the equality of
electromagnetic and gravitational interaction strengths at the
Planck scale. Using (\ref{e10}) in (\ref{ex}), we can deduce that
\begin{equation}
Gm^2 \sim \epsilon^2/10^{40}\label{ey}
\end{equation}
The relation (\ref{ey}) has been well known, though as an empirical
relation expressing the relative strengths of the gravitational and
electromagnetic strengths. Here we have deduced it on the basis of
our theory.\\
The Compton scale (\ref{4ea2}) comes as a Quantum Mechanical effect,
within which we have zitterbewegung effects and a breakdown of
causal Physics as emphasized in the literature \cite{diracpqm}.
Indeed Dirac had noticed this aspect in connection with two
difficulties with his electron equation. Firstly the speed of the
electron turns out to be the velocity of light. Secondly the
position coordinates become complex or non Hermitian. His
explanation was that in Quantum Theory we cannot go down to
arbitrarily small spacetime intervals, for the Heisenberg
Uncertainty Principle would then imply arbitrarily large momenta and
energies. So Quantum Mechanical measurements are actually an average
over intervals of the order of
the Compton scale.\\
Weinberg also noticed the non physical aspect of the Compton scale
\cite{weinberggc}. Starting with the usual light cone of Special
Relativity and the inversion of the time order of events, he goes on
to add, and we quote,\\
``Although the relativity of temporal order raises no problems for
classical physics, it plays a profound role in quantum theories. The
uncertainty principle tells us that when we specify that a particle
is at position $x_1$ at time $t_1$, we cannot also define its
velocity precisely. In consequence there is a certain chance of a
particle getting from $x_1$ to $x_2$ even if $x_1 - x_2$ is
spacelike, that is, $| x_1 - x_2 | > |x_1^0 - x_2^0|$. To be more
precise, the probability of a particle reaching $x_2$ if it starts
at $x_1$ is nonnegligible as long as
$$(x_1 - x_2)^2 - (x_1^0 - x_2^0)^2 \leq \frac{\hbar^2}{m^2}$$
where $\hbar$ is Planck's constant (divided by $2\pi$) and $m$ is
the particle mass."
\section{Discussion}
1. It may be mentioned that the Compton wavelength in the context of
the background vacuum energy or Dark Energy, as we saw in
(\ref{4e1}) has the following important property \cite{bgsarxiv}:
The Coulomb self energy which is proportional to $1/a$ where $a$ in
which case is the Compton wavelength, exactly balances the vacuum
energy, thus providing a stable configuration. In the old theory on
the other hand, this was a major inconsistency-- neither could the
length $a$ could be non zero nor could $a \to 0$ as the self energy
would then diverge \cite{rohr}.\\
2. We can get a further insight into the array of Planck
oscillators, following the earlier argument of Random fluctuational
creation of such oscillators from the background Dark Energy,
alluded to in the cosmological model at the beginning of Section 1.
According to this Planck oscillators are randomly created and
destroyed in the Dark Energy background. However, mimicking the
Random Walk, there would be a nett creation of $\sqrt{n}$ Planck
oscillators out of $n$ total fluctuations in a time interval $t_P$,
the Planck time, which in any case is very small and is nearly a
continuum. So we have
\begin{equation}
\frac{dn}{dt} \approx \frac{\sqrt{n}}{t_P}\label{ez}
\end{equation}
Integrating (\ref{ez}) we get
$$t_\pi = \sqrt{n} t_P \, \mbox{or}\, l_\pi = \sqrt{n} l_P$$
which is just equation (\ref{4De1d}), giving $n \sim 10^{40},$ for
the Compton time of an elementary particle like the pion.\\
On the other hand integrating up to the age of the universe
$$T \sim 10^{17},$$
we get
$$T = \sqrt{N} t_P,$$
giving $N \sim 10^{120}$. This gives a rationale for the values of
$n$ and $N$ given above.\\
3. We can consider the above scenario from yet another point of
view, that of Quantum Statistical Mechanics. Here also, in the
spirit of randomness, the state can be written as \cite{cu,huang}
\begin{equation}
\psi = \sum_{n} c_n \phi_n,\label{e5}
\end{equation}
in terms of basic states $\phi_n$ representing a Planck oscillator
with energy $E_n$ it is known that (\ref{e5}) can be rewritten as
\begin{equation}
\psi = \sum_{n} b_n \bar{\phi}_n\label{e6}
\end{equation}
where $|b_n|^2 = 1$ if $E<E_n<E+\Delta$, and $= 0$ otherwise under
the assumption
\begin{equation}
\overline{(c_n,c_m)} = 0, n \ne m\label{e7}
\end{equation}
(Infact $n$ could stand for not a single state but for a set of
states $n_\imath$ and so also $m$). Here the bar denotes a time
average over a suitable interval. This is the well known Random
Phase Axiom and arises due to the total randomness amongst the
phases $c_n$. Also the expectation value of any operator $O$ is
given by
\begin{equation}
<O> = \sum_{n} |b_n|^2 (\bar{\phi}_n, O \bar{\phi}_n)/ \sum_{n}
|b_n|^2\label{e8}
\end{equation}
(\ref{e6}) and (\ref{e8}) show that we have incoherent states
$\bar{\phi}_1, \bar{\phi}_2,etc$ once averages over time intervals
for the phases $c_n$ in (\ref{e7})vanish owing to their relative
randomness. Here while the state $\phi_N$ in (\ref{e5}) represent
the Planck oscillators, the states $\bar{\phi}_1,\bar{\phi}_2 etc.$
in (\ref{e6}) or (\ref{e8}) represent the energy states of
elementary particles, which averageover $10^{40}$ Planck oscillator
states.
\section{Conclusion}
We have thus argued from different independent points of view that
an underpinning of Planck oscillators in a background of Dark Energy
explains in a thermodynamic sense, why the universe settles at the
real life elementary particle scale rather than the Planck scale.

\end{document}